\documentclass[11pt,eqs]{article}
\usepackage{latexsym}
\usepackage{amsmath}

\makeatletter
\newcommand\xleftrightarrow[2][]{%
  \ext@arrow 9999{\longleftrightarrowfill@}{#1}{#2}}
\newcommand\longleftrightarrowfill@{%
  \arrowfill@\leftarrow\relbar\rightarrow}
\makeatother

\title{
Directly from $H$-flux to the family of three nonlocal $R$-flux theories 
\thanks{Work supported in part by
the Serbian Ministry of Education, Science and Technological Development, under contract No. 171031. 
I also want to thank to Prof. Dr. Branislav Sazdovi\'c and Dr. Ljubica Davidovi\'c from Institute of Physics Belgrade
for useful discussions.}}
\author{B. Nikoli\'c, D. Obri\'c \thanks{email: bnikolic, dobric@ipb.ac.rs}\\{\it Institute of Physics Belgrade, University of Belgrade, Pregrevica 118, Serbia} }

\begin{document}

\maketitle
\begin{abstract}
In this article we consider T-dualization of the 3D closed bosonic string in the weakly curved background - constant metric and Kalb-Ramond field with one non-zero component, $B_{xy}=Hz$, 
where field strength $H$ 
is infinitesimal. 
We use standard and generalized Buscher T-dualization procedure and make T-dualization starting from coordinate $z$, via $y$ and finally along $x$ coordinate. All three theories are {\it nonlocal}, 
because variable $\Delta V$, defined as line integral, appears as an argument of background fields. After the first T-dualization we obtain commutative and associative theory, while after 
we T-dualize along $y$, we get, $\kappa$-Minkowski-like, noncommutative and associative theory. At the end of this T-dualization chain we come to the theory which is both noncommutative and nonassociative. 
The form of the final T-dual action does not depend on the order of T-dualization while noncommutativity and nonassociativity relations could be obtained from those in the  $x\to y\to z$ case by
replacing $H\to - H$.
\end{abstract}

\section{Introduction}
\setcounter{equation}{0}

Noncommutativity of coordinates has come into focus of physics about hundred years ago when the problem with infinite value of physical quantities occurred. 
The solution was proposed by Heisenberg in the form of noncommutative coordinates. But after developing of renormalization procedure coordinate noncommutativity 
was forgotten as a tool for cancelling of infinities.

Commuting of coordinates means that there is no minimal possible length in Nature i.e. that we can measure the position of particle with infinite precision. The return of noncommutativity into physics starts 
with
the article of Hartland Snyder \cite{snyder}. Usually we treat space-time as continuum but Snyder showed that there is Lorentz invariant discrete space-time. Consequently, this means that 
commutator of coordinates is nonzero, and noncommutativity parameter dictates the scale at which noncommutativity exists.

In the paper \cite{SW} existence of noncommutative manifold was shown using propagators in open bosonic string theory with constant metric and constant Kalb-Ramond field. 
This result is proven in many articles \cite{gbnc} after that but using different mathematical methods. Obtained noncommutativity with constant noncommutativity parameter is known in 
literature as canonical noncommutativity. Consequently, canonical noncommutativity implies that theory is still associative one.

One of the first application of canonical noncommutativity was in Yang-Mills (YM) theories \cite{ncym}. 
Noncommutative YM theories are constructed and their renormalisability properties are analyzed. It turned out that some processes forbidden in commutative YM are 
allowed in noncommutative YM theories. Consequently, cross sections for those decays and processes are calculated \cite{expnc}. Such predictions offer the possibility of indirect check 
of idea of noncommutativity.

The next type of noncommuatativity which is considered in literature is Lie-algebraic one, which means that commutator of two coordinates is proportional to the coordinate. The $\kappa$-Minkowski space-time is an
example of this kind of noncommutativity and it is considered in various contexts \cite{kappa}. The $\kappa$-Minkowski space is noncommutative but it is easy to check that is associative one. But, in general, 
if the commutator of the coordinates is proportional to the some linear combination of coordinates, then the space is nonassociative because jacobiator and associator are nonzero. For example, such spaces are closely related to
the $L_{\infty}$ algebra \cite{blumh}.

The mathematical framework for T-dualization is standard Buscher procedure \cite{B}. 
It consists of the localization of the shift symmetry and adding a term with Lagrange multiplier in order to make gauge fields 
unphysical degrees of freedom. Also there is an improvement of standard Buscher procedure developed and applied in Refs.\cite{DS1,DNS2,DNS,DNS3}, 
generalized Buscher procedure. In the application of the generalized procedure of T-dualization there is one additional step with respect to the standard one. 
We introduce invariant coordinate in order to localize shift symmetry in the coordinate dependent backgrounds.

The first articles addressing the subject of coordinate dependent backgrounds  appear in the last ten years \cite{Lust,ALLP,ALLP2,nemci,L,nongeo1,nongeo2}. 
A 3-torus with constant metric and Kalb-Ramond field with just one nonzero component, $B_{xy}=Hz$, was considered within standard Buscher procedure \cite{ALLP}. 
Authors made two successive T-dualzation along isometry directions $x$ and $y$, and, using nontrivial winding conditions, obtained noncommutativity with parameter proportional
to field strength $H$ and winding number $N_3$.

Using generalized T-duality procedure \cite{DNS,bndo} we obtained coordinate dependent noncommutativity and, consequently, nonassociativity. 
Also it is shown that final theory is nonlocal. In Ref.\cite{DNS} the bosonic string is considered in the weakly curved background - constant metric and linearly coordinate dependent Kalb-Ramond field with infinitesimal field strength, while in \cite{bndo} we consider the same model as in \cite{ALLP}, but T-dualizing along all three directions and imposing trivial winding conditions. Obtained nonlocality comes from the coordinate dependent background, or more precisely, from invariant coordinates. At the end of T-dualization procedure background fields depend on $\Delta V$, defined as line integral. Nonlocality has been become very important issue in the quantum mechanical considerations \cite{nonlocal}.

In this article we will deal with closed bosonic string propagating in the constant metric and linear dependent Kalb-Ramond field with $B_{xy}=Hz$, the same background as in \cite{ALLP,bndo}. But our goal here is to examine the influence of order of T-dualizations. In Ref.\cite{bndo} we T-dualize first along isometry directions, first along $x$ and then along $y$, and at the end, along direction $z$. The first T-dualization produces configuration known as twisted torus which is commutative, and it is globally and locally well defined. After second T-dualization we obtained nongeometric theory with $Q$ flux which is still locally well defined and it is commutative. The final T-dualization along $z$ direction produces nonlocal theory which is noncommutative and nonassociative one. This line of T-dualizations we will call $xyz$ one.

But what it will happen, if we change the order of T-dualizations, regrading (non)lo\-ca\-li\-ty issue as well as (non)commutativity and (non)associativity? 
It is quite obvious that nothing will be changed if we T-dualize along line $yxz$, because the first two directions, which are T-dualized, are isometry ones. 
Some nontrivial issues could be expected if T-dualize first along $z$ direction. In this article we will present T-dualization of the model from \cite{ALLP,bndo} along the T-dualization line $zyx$. 
After every step of T-dualization we will rewrite the T-dual transformation law in canonical form using the expressions for canonical momenta of the initial theory. 
Also we will check whether the obtained theory is commutative or not and, consequently, we will see whether it is associative or not.

The fact which is quite sure is that all three theories which we will obtain from the T-dualization line $zyx$ are {\it nonlocal}. 
The explanation comes from the fact that background field $B_{\mu\nu}$ is $z$ dependent and according to the generalized T-dualization procedure, after T-dualization along $z$, we obtain quantity $\Delta V$ 
which is defined as line integral. Consequently, the theory is nonlocal. But because $y$ and $x$ T-dualizations do not affect $\Delta V$, all three theories obtained in $zyx$ T-dualization line 
are nonlocal. That is a difference with respect to the $xyz$ T-dualization line considered in \cite{bndo}. 

The interesting thing is that transformation laws can be obtained from the corresponding ones in \cite{bndo} by replacing $H\to - H$, 
but because in this article we T-dualize in the opposite direction, that produces theories of the different commutative and associative features with respect to \cite{bndo}. 
After first T-dualization we get commutative and associative theory which is the same as in $xyz$ case from \cite{bndo}. But the second T-dualization here produces {\it noncommutative} and associative theory 
of $\kappa$-Minkowski type.
That is different with respect to the $xyz$ case, where second theory in the line is both commutative and associative. At the end we obtain the same theory as in \cite{bndo} which is nonassociative and noncommutative. The noncommutativity and nonassociativity
parameters have one additional "$-$" sign comparing with the corresponding ones in \cite{bndo}.
In this article as well as in \cite{bndo}, we impose trivial winding conditions which means $x^\mu(\sigma+2\pi)=x^\mu(\sigma)+2\pi N^\mu$, where $N^\mu$ is a winding number.

At the end we comment some quantum aspects of the problem and add two appendices. The first one contains conventions regarding light-cone coordinates, while the second one is related to the mathematical details concerning derivation of two kinds of Poisson
brackets appearing in the article.


\section{Bosonic string action and choice of background fields}
\setcounter{equation}{0}

In this section we will introduce the action for bosonic string propagating in 3D space with constant metric and Kalb-Ramond field which single component is different from zero, 
$B_{xy}=Hz$. This model is well known in literature as torus with $H$-flux. Since we are working with the same model as in \cite{ALLP,bndo}, for completeness we will repeat most of the steps from introductory 
part in the \cite{bndo}.

The closed bosonic string which propagates in the
presence of the space-time metric $G_{\mu\nu}(x)$, Kalb-Ramond
field $B_{\mu\nu}(x)$, and dilaton field
$\Phi(x)$ is described by action \cite{S}
\begin{equation}\label{eq:action2}
S = \kappa  \int_\Sigma  d^2 \xi  \sqrt{-g}  \left\{  \left[  \frac{1}{2}g^{\alpha\beta}G_{\mu\nu}(x) +\frac{\varepsilon^{\alpha\beta}}{\sqrt{-g}}  B_{\mu\nu}(x) \right] \partial_\alpha x^\mu
\partial_\beta x^\nu +  \Phi (x) R^{(2)}  \right\} \,  ,
\end{equation}
where world-sheet surface $\Sigma$ is parameterized by $\xi^\alpha=(\tau\, ,\sigma)$ [$(\alpha=0\, ,1)$,
$\sigma\in(0\, ,\pi)$], while $x^\mu$ ($\mu=0,1,2,\dots,D-1$) are space-time coordinates. Intrinsic world sheet metric is denoted by $g_{\alpha\beta}$, and the corresponding scalar curvature with
$R^{(2)}$.

Conformal symmetry on the quantum level is not preserved for any choice of background fields. If we want to keep conformal symmetry on the quantum level, background fields must obey the space-time field equations \cite{CFMP}
\begin{equation}\label{eq:betaG}
\beta^G_{\mu \nu} \equiv  R_{\mu \nu} - \frac{1}{4} B_{\mu \rho
\sigma} B_{\nu}{}^{\rho \sigma} +2 D_\mu a_\nu =0   \,  ,
\end{equation}
\begin{equation}\label{eq:betaB}
\beta^B_{\mu \nu} \equiv  D_\rho B^\rho{}_{\mu \nu} -2 a_\rho
 B^\rho{}_{\mu \nu} =0  \,  ,
\end{equation}
\begin{equation}\label{eq:betaFi}
\beta^\Phi \equiv 2\pi \kappa\frac{D-26}{6}-R - \frac{1}{24} B_{\mu \rho \sigma}
B^{\mu \rho \sigma} -  D_\mu a^\mu + 4 a^2 =c \,  ,
\end{equation}
where $c$ is an arbitrary constant. From
\begin{equation}
D^\nu \beta_{\nu\mu}^G+\partial_\mu \beta^\Phi=0\, ,
\end{equation}
it follows that third beta function, $\beta^\Phi$, is equal to an arbitrary constant.
Here $R_{\mu \nu}$ and $D_\mu$ are Ricci tensor
and covariant derivative with respect to the space-time metric $G_{\mu\nu}$. Field strength for Kalb-Ramond field $B_{\mu\nu}$ and dilaton gradient
are defined as
\begin{equation}
B_{\mu \nu \rho}=\partial_\mu B_{\nu\rho}+\partial_\nu
B_{\rho\mu}+\partial_\rho B_{\mu\nu}\, ,\quad a_\mu=\partial_\mu
\Phi\, .
\end{equation}

One of the solutions of these equations which is important for us here is the solution where some background fields are coordinate dependent.
Let us choose Kalb-Ramond field to be linearly coordinate dependent and dilaton field to be constant.
The equation (\ref{eq:betaG}) turns into
\begin{equation}
R_{\mu \nu} - \frac{1}{4} B_{\mu \rho
\sigma} B_{\nu}{}^{\rho \sigma}=0\, .
\end{equation}
If we assume that field strength is infinitesimal, then we take $G_{\mu\nu}$ to be constant in approximation linear in $B_{\mu\nu\rho}$. Consequently, 
 the third equation (\ref{eq:betaFi}) is of the form
\begin{equation}
2\pi \kappa\frac{D-26}{6}=c\, .
\end{equation}
The constant $c$ is arbitrary, and fixing its value at $c=-\frac{23\pi\kappa}{3}$, we obtain $D=3$, dimension of the space in which we will work further.  

The choice of background fields in the case we will consider is
\begin{equation}
G_{\mu\nu}=\left(
\begin{array}{ccc}
R_1^2 & 0 & 0\\
0 & R_2^2 & 0\\
0 & 0 & R_3^2
\end{array}\right)\, ,\quad B_{\mu\nu}=\left(
\begin{array}{ccc}
0 & Hz &0\\
-Hz & 0 &0\\
0 & 0 & 0
\end{array}\right)\, ,
\end{equation}
where $R_\mu (\mu=1,2,3)$ are radii of the compact dimensions. In terms of radii, the imposed condition that $H$ is infinitesimal, can be rewritten as
\begin{equation}
(\frac{H}{R_1 R_2 R_3})^2=0\, .
\end{equation}
Physically, infinitesimality of $H$ means that we work with  sufficiently large torus (diluted flux approximation). If we rescale the coordinates
\begin{equation}
x^\mu\longmapsto x'^\mu=R_\mu x^\mu\, ,
\end{equation}
where indices on the right hand-side of equation are not summed,
the form of the metric simplifies
\begin{equation}
G_{\mu\nu}=\left(
\begin{array}{ccc}
1 & 0 & 0\\
0 & 1 & 0\\
0 & 0 & 1
\end{array}\right)\, .
\end{equation}

Taking all assumption into consideration, the action is of the form
\begin{eqnarray}\label{eq:dejstvo}
S&=&\kappa\int_\Sigma d^2\xi \partial_+ x^\mu \Pi_{+\mu\nu}\partial_- x^\nu\\ &=&\kappa\int_\Sigma d^2\xi \left[\frac{1}{2}\left(\partial_+ x \partial_- x+\partial_+ y \partial_-y +\partial_+z \partial_-z\right)+\partial_+x Hz \partial_-y-\partial_+y Hz \partial_-x\right]\nonumber\, ,
\end{eqnarray}
where $\partial_\pm=\partial_\tau\pm \partial_\sigma$ is world-sheet derivative with respect to the light-cone coordinates $\xi^\pm=\frac{1}{2}(\tau\pm\sigma)$, $\Pi_{\pm \mu\nu}=B_{\mu\nu}\pm \frac{1}{2}G_{\mu\nu}$ and 
\begin{equation}
x^\mu=\left(
\begin{array}{c}
x\\y\\z
\end{array}\right)\, .
\end{equation}
T-dualization of dilaton is done within quantum formalism and here it will not be presented.

\section{Family of three $R$ flux non-local theories}
\setcounter{equation}{0}

In this section we will perform T-dualization of closed bosonic string equipped by $H$-flux torus background fields, one direction at time. 
T-dualization procedure will go along $zyx$ line. We will show that all three theories are nonlocal with $R$-flux. 
Also we will find expressions connecting initial and T-dual variables, so called T-dual transformation laws. Using transformation laws in canonical form, 
we will check after every step whether obtained theory is (non)commutative and/or (non)associative.

\subsection{T-dualization along $z$ direction - shortcut to $R$-flux}

Unlike the cases considered in \cite{ALLP,bndo}, where T-dualization drives along $xyz$ line, let us do that in opposite direction and perform generalized T-dualization \cite{DS1} of 
action (\ref{eq:dejstvo}) along $z$ direction. 

\subsubsection{T-dualization procedure}

It looks like that this direction is not isometry one. But we can show that it can be treated like isometry direction. Let us consider the global transformation
\begin{equation}
\delta x^\mu=\lambda^\mu\, ,
\end{equation}
and vary the action with respect to this transformation
\begin{equation}\label{eq:povc}
\delta S=\frac{\kappa}{3}B_{\mu\nu\rho}\lambda^\rho\int_{\Sigma}d^2\xi\partial_+x^\mu\partial_-x^\nu=\frac{2k}{3}B_{\mu\nu\rho}\lambda^\rho\epsilon^{\alpha\beta}\int_{\Sigma}d^2\xi[\partial_\alpha(x^\mu\partial_\beta x^\nu)-x^\mu(\partial_\alpha\partial_\beta x^\nu)]\, . 
\end{equation}
The second term vanishes as a consequence of contraction of antisymmetric ($\epsilon^{\alpha\beta}$) and symmetric ($\partial_\alpha\partial_\beta$) tensors, 
while the first one, surface term, survives, and it is, in
general, different from zero. But, the expression $\delta S$ is an topological invariant,
so it vanishes if the map from the world-sheet to $D$-dimensional space-time is
topologically trivial. Essentially, infinitesimal field strength $H$ does not affect
the vanishing of the surface term.

There is one more explanation of vanishing of this surface term. It is more
technical and adjusted to the approximation we used in this article which essence is the explanation in paragraph above. Because
we work in the approximation up to the linear terms in $H$, $x^\mu$ satisfies equation
of motion for constant $G_{\mu\nu}$ and $B_{\mu\nu}$, $\partial_+\partial_- x^\mu=0$, which solution is well known in
literature. If the winding number is equal to zero, it holds $x^\mu(2\pi+\sigma)=x^\mu(\sigma)$, and since the configuration in the initial $\tau_i$ and final moment $\tau_f$ is fixed, the surface term vanishes.

So, in the weakly curved background case ($H$-flux torus background is such like that), $z$ direction is an isometry one.
Localization of the shift symmetry of the action (\ref{eq:dejstvo}) along $z$ starts with introducing the covariant derivative
\begin{equation}
\partial_\pm z \longrightarrow D_\pm z=\partial_\pm z+v_\pm\, ,
\end{equation}
where $v_\pm$ is a gauge field. In order to make gauge fields unphysical ones, we introduce term with Lagrange multiplier
\begin{equation}
S_{add}=\frac{\kappa}{2}\int_{\Sigma} d^2\xi y_3 (\partial_+ v_--\partial_- v_+)\, .
\end{equation}
These two steps are the part of the standard Buscher procedure. Because of coordinate dependent background field $B_{\mu\nu}$, generalized T-dualization procedure has an additional step, introducing of an invariant coordinate
\begin{equation}
z^{inv}=\int_P d\xi^\alpha D_{\alpha}z=\int_P d\xi^+ D_+ z+\int_P d\xi^- D_- z= z(\xi)-z(\xi_0)+\Delta V\, ,
\end{equation}
where
\begin{equation}\label{eq:deltav}
\Delta V=\int_{P}d\xi^\alpha v_\alpha=\int_P (d\xi^+ v_++d\xi^- v_-)\, .
\end{equation}
The form of the action is now
\begin{eqnarray}
\bar{S}&=&\kappa\int_{\Sigma}d^2\xi\left[Hz^{inv}(\partial_+x\partial_- y-\partial_+ y\partial_- x)
+\frac{1}{2}(\partial_+ x\partial_- x+\partial_+ y\partial_- y+D_+ zD_-z) \right. \nonumber \\
&+&\left.\frac{1}{2}\; y_3 (\partial_+ v_--\partial_- v_+)\right]\, .
\end{eqnarray}
Fixing the gauge, $z(\xi)=z(\xi_0)$, we get gauged fixed action in the form
\begin{eqnarray}\label{eq:gfa1}
{S}_{fix}&=&\kappa\int_{\Sigma}d^2\xi\left[H\Delta V
(\partial_+ x\partial_- y-\partial_+ y\partial_-x)
+\frac{1}{2}(\partial_+ x\partial_-x+\partial_+y\partial_-y+v_+v_-)\right. \nonumber\\ 
&+&\left.\frac{1}{2}y_3 (\partial_+ v_--\partial_- v_+)\right]\, .
\end{eqnarray} 

The equation of motion for Lagrange multiplier $y_3$ obtained from above action (\ref{eq:gfa1}) produces 
\begin{equation}\label{eq:eomv3}
\partial_+ v_--\partial_- v_+=0\Longrightarrow v_\pm=\partial_\pm z\, ,
\end{equation}
which drives us back to the initial action (\ref{eq:dejstvo}). On the other side, if we found equations of motion for gauge fields $v_\pm$, we get
\begin{equation}\label{eq:izrazv1}
v_\pm = \pm \partial_\pm y_3-2\beta^{\mp}\, ,
\end{equation}
where $\beta^\pm$ functions are defined as
\begin{equation}
\beta^\pm=\mp\frac{1}{2}H(x\partial_\mp y-y\partial_\mp x)\, .
\end{equation}
The $\beta^\pm$ functions stem from the variation of the term containing $\Delta V$. 
The derivation of beta functions $\beta^\pm$ is based on the relation
$\partial_\pm\Delta V=v_\pm$. In the derivation
of the beta functions there is one nontrivial technical point and that is
vanishing of the surface term after one partial integration. That surface term is of
the same form as in Eq.(\ref{eq:povc}), so the same reasons for surface term vanishing
hold here.
Mathematical details regarding derivation of $\beta^\pm$ functions can be found in Refs.\cite{DS1,DNS2,DNS,DNS3,bndo}.

Inserting the relations (\ref{eq:izrazv1}) into the gauge fixed action, keeping linear terms in $H$, we obtain the T-dual action
\begin{equation}\label{eq:sz}
{}_{z} S=\kappa \int_\Sigma d^2 \xi \partial_+ {}_{z}X^\mu {}_{z} \Pi_{+\mu\nu} \partial_- {}_{z} X^\nu\, ,
\end{equation}
where 
\begin{equation}
{}_{z}X^\mu=\left(
\begin{array}{c}
x \\ y \\ y_3                    
\end{array}\right)\, ,\quad {}_{z}\Pi_{+\mu\nu}={}_{z} B_{\mu\nu}+\frac{1}{2}{}_{z} G_{\mu\nu}\, ,
\end{equation}
\begin{equation}
{}_{z}B_{\mu\nu}=\left(
\begin{array}{ccc}
0 & H\Delta V & 0\\
-H\Delta V & 0 & 0\\
0 & 0 & 0
\end{array}\right)\, ,\quad {}_{z}G_{\mu\nu}=\left(
\begin{array}{ccc}
1 & 0 & 0\\
0 & 1 & 0\\
0 & 0 & 1
\end{array}\right)\, .
\end{equation}

Let us note that presence of $\Delta V$, defined as line integral,
represents the source of nonlocality of the T-dual theory.

\subsubsection{T-dual transformation law}

Combining the equations of motion for Lagrange multiplier (\ref{eq:eomv3}) and for gauge fields (\ref{eq:izrazv1}), we obtain T-dual transformation laws
\begin{equation}\label{eq:zakon1}
\partial_\pm z\cong \pm \partial_\pm y_3\mp H(x\partial_\pm y-y\partial_\pm x)\, ,                                                                                                                                                        
\end{equation}
where $\cong$ is used here to mark T-dual relation.
Momentum of the initial theory (\ref{eq:dejstvo}) canonically conjugated to the coordinate $z$ is of the form
\begin{equation}
\pi_z=\frac{\partial{\mathcal L}}{\partial \dot z}=\kappa \dot z\, ,
\end{equation}
where $\mathcal L$ is a Lagrangian density defined as $S=\int_\Sigma d^2\xi \mathcal L$. Calculating $\dot z$ using T-dual transformation law (\ref{eq:zakon1}), we get the T-dual transformation law in canonical form
\begin{equation}\label{eq:y3}
y_3'\cong \frac{1}{\kappa} \pi_z +H(xy'-yx')\, ,
\end{equation}
which is of the same form as in the $xyz$ case.

In all further expressions we will keep the symbol $\Delta V$, but we must have in mind that we used equations of motion for Lagrange multipliers 
(\ref{eq:eomv3}) at the end of T-dulization procedure along $z$ coordinate, so, having in mind (\ref{eq:deltav}) and (\ref{eq:zakon1}), we get
\begin{equation}
\Delta V=\Delta z\cong \int d\xi^+ \partial_+ y_3-\int d\xi^- \partial_- y_3\equiv \tilde y_3\, .
\end{equation}
The variable $\Delta V$ is multiplied by infinitesimal field strength $H$, so, in the above expression we used $\partial_\pm z\cong \pm \partial_\pm y_3$, as a consequence of diluted flux approximation.
                                                                                                                                            
\subsubsection{(Non)commutativity and (non)associativity}

The initial theory is geometric one and its variables satisfy the standard Poisson algebra
\begin{equation}\label{eq:spa}
\{x^\mu(\sigma),x^\nu(\bar\sigma)\}=\{\pi_\mu(\sigma),\pi_\nu(\bar\sigma)\}=0\, ,\quad \{x^\mu,\pi_\nu(\bar\sigma)\}=\delta^\mu{}_\nu \delta(\sigma-\bar\sigma)\, ,
\end{equation}
where $x^\mu$ are the coordinates of the initial theory, while $\pi_\mu$ are their canonically conjugated momenta. Using expression (\ref{eq:y3}) and standard Poisson algebra (\ref{eq:spa}), we obtain that coordinates of the theory obtained after one T-dualization, ${}_z X^\mu$, are commutative. Consequently, Jacobiator is equal to zero, which means that theory is associative. 

Summarizing this first step of T-dualization, obtained theory is {\bf commutative and associative nonlocal R-flux theory}. 
Comparing with the results of the Ref.\cite{bndo} after first T-dualization, qualitatively we obtain the same result, but with the essential difference that here obtained theory is nonlocal R-flux theory  unlike that in \cite{bndo} which is 
geometrical one, locally and globally
well defined.

\subsection{Step 2 - T-dualization along $y$ direction}

Our starting point is the action given in Eq.(\ref{eq:sz}). The background fields are independent of $y$, so, we apply standard Buscher procedure. This means that, unlike the previous case, we perform just first two steps in T-dualization procedure and skip the third one - introducing of invariant coordinate. The T-dualization procedure is already presented, so, we will skip explaining procedure steps further.

\subsubsection{T-dualization procedure}

The gauge fixed action is of the form
\begin{eqnarray}\label{eq:sfix2}
S_{fix}&=&\kappa \int_\Sigma d^2 \xi \left[\frac{1}{2}\left(\partial_+ x\partial_- x+v_+ v_-+\partial_+ y_3 \partial_- y_3\right)+H\Delta V(v_-\partial_+ x - v_+\partial_- x)\right]+\nonumber\\
&+&\frac{\kappa}{2}\int_\Sigma d^2\xi y_2 (\partial_+ v_--\partial_- v_+)\, .
\end{eqnarray}

Varying with respect to the Lagrange multiplier $y_2$ we get
\begin{equation}\label{eq:v2}
v_\pm=\partial_\pm y\, ,
\end{equation}
while the equations of motion for gauge fields are
\begin{equation}\label{eq:vpm2}
v_\pm=\pm \partial_\pm y_2 \mp 2H\Delta V \partial_\pm x\, .
\end{equation}
Inserting the expression for gauge fields (\ref{eq:vpm2}) into gauge fixed action (\ref{eq:sfix2}), we obtain the T-dual action
\begin{equation}
{}_{zy} S=\kappa \int_\Sigma d^2\xi \;\partial_+\; {}_{zy} X^\mu \; {}_{zy}\Pi_{+\mu\nu}\; \partial_-\; {}_{zy} X^\nu\, ,
\end{equation}
where 
\begin{equation}
{}_{zy} X^\mu=\left(
\begin{array}{c}
x\\
y_2\\
y_3
\end{array}\right)\, ,\quad {}_{zy}\Pi_{+\mu\nu}={}_{zy} B_{\mu\nu}+\frac{1}{2}\;{}_{zy}G_{\mu\nu}\, ,
\end{equation}
\begin{equation}
{}_{zy} B_{\mu\nu}=0\, ,\quad {}_{zy} G_{\mu\nu}=\left(
\begin{array}{ccc}
1 & -2H\Delta V & 0\\
-2H\Delta V & 1 & 0\\
0 & 0 & 1
\end{array}
\right)\, .
\end{equation}
Let us note that after two T-dualizations in the $xyz$ case in \cite{bndo} we also obtained that T-dual Kalb-Ramond field is zero. 

\subsubsection{T-dual transformation law}

Combining equations of motion (\ref{eq:v2}) and (\ref{eq:vpm2}) we get the corresponding transformation law
\begin{equation}\label{eq:tl2}
\partial_\pm y\cong \pm \partial_\pm y_2\mp 2H\Delta V \partial_\pm x\, .
\end{equation}
\
Let us now prescribe the transformation law in canonical form. The momentum 
canonically conjugated to the initial coordinate $y$ is obtained by variation of the initial action (\ref{eq:dejstvo}) with respect to the $\dot y$ and it is of the form
\begin{equation}
\pi_y=\kappa(\dot y+2Hz x')\, ,
\end{equation}
while from transformation law (\ref{eq:tl2}) we have
\begin{equation}
\dot y\cong y_2'-2H\Delta V x'\, .
\end{equation}
Combining last two equations and using the fact that, in the approximation linear in $H$, $\Delta V$ and $z$ are T-dual to each other, we get
\begin{equation}\label{eq:y2}
y_2'\cong \frac{1}{\kappa}\pi_y\, .
\end{equation}
As we see the transformation law is the same as in the $xyz$ case.

\subsubsection{(Non)commutativity and (non)associativity}

In this paragraph we will calculate Poisson brackets of the coordinates ${}_{zy}X^\mu$ using transformation laws in canonical form given by Eqs.(\ref{eq:y3}) and (\ref{eq:y2}).

With the help of the standard Poisson algebra (\ref{eq:spa}) and instructions from Appendix B, it is easy to see that
\begin{equation}
\{x(\sigma),x(\bar\sigma)\}=\{y_2(\sigma),y_2(\bar\sigma)\}=\{y_3(\sigma),y_3(\bar\sigma)\}=\{x(\sigma),y_2(\bar\sigma)\}=\{x(\sigma),y_3(\bar\sigma)\}=0\, .
\end{equation}
The only non-zero Poisson bracket is
\begin{equation}\label{eq:y2y3}
\{y_2'(\sigma),y_3'(\bar\sigma)\}\cong \frac{H}{\kappa}\left[2x'(\sigma)\delta(\sigma-\bar\sigma)+x(\sigma)\delta'(\sigma-\bar\sigma)\right]\, ,
\end{equation}
where $\delta'\equiv \partial_\sigma\delta(\sigma-\bar\sigma)$. This result is obtained by straightforward calculation using 
T-dual transformation laws, (\ref{eq:y3}) and (\ref{eq:y2}), and standard Poisson algebra (\ref{eq:spa}). The relation (\ref{eq:y2y3}) is of the form (\ref{eq:sigmapoisson}), where $A'(\sigma)=y_2'(\sigma) $, $B'(\bar\sigma)=y_3'(\bar\sigma)$, $U'(\sigma)=\frac{H}{\kappa}2x'(\sigma)$ and $V(\sigma)=\frac{H}{\kappa}x(\sigma)$. With these substitutions in mind, we have that final expression is of the form (\ref{eq:tdpoisson})
\begin{equation}\label{eq:y2y3p}
\{y_2(\sigma),y_3(\bar\sigma)\}\cong-\frac{H}{\kappa}\left[2x(\sigma)-x(\bar\sigma)\right]\theta(\sigma-\bar\sigma)\, .
\end{equation}
For $\sigma\to\sigma+2\pi$ and $\bar\sigma\to \sigma$ we have
\begin{equation}
\{y_2(\sigma+2\pi),y_3(\sigma)\}\cong-\frac{H}{\kappa}\left[x(\sigma)+4\pi N_x\right]\, ,
\end{equation}
because $\theta(2\pi)=1$ (\ref{eq:fdelt}), while $N_x$ is winding number for $x$ coordinate
\begin{equation}\label{eq:winding}
x(\sigma+2\pi)-x(\sigma)=2\pi N_x\, .
\end{equation}
As we can see the noncommutativity relation (\ref{eq:y2y3p}) is of $\kappa$-Minkowski type. It is straightforward to see that
\begin{equation}
\{x(\sigma_1),\{y_2(\sigma_2),y_3(\sigma_3)\}\}+\{y_2(\sigma_2),\{y_3(\sigma_3),x(\sigma_1)\}\}+\{y_3(\sigma_3),\{x(\sigma_1),y_2(\sigma_2)\}\}\cong 0\, .
\end{equation}
Because the Jacobiator is zero, we conclude that this R-flux theory is {\bf noncommutative} and {\bf associative} one.

\subsection{Step 3 - T-dualization along $x$ direction}

In this subsection we will finish T-dualization procedure not repeating the mathematical details, but giving just the important equations and results.

The gauge fixed action is given by the following equation
\begin{eqnarray}
S_{fix}&=&\kappa\int_\Sigma d^2\xi \left[\frac{1}{2}\left(v_+ v_-+\partial_+ y_2\partial_- y_2+\partial_+ y_3\partial_- y_3\right)-H\Delta V(v_+ \partial_-y_2+\partial_+y_2 \;v_-)\right]\nonumber \\
&+&\frac{\kappa}{2}\int_\Sigma d^2\xi y_1(\partial_+ v_--\partial_- v_+)\, .
\end{eqnarray}
The equations of motion for Lagrange multiplier produces
\begin{equation}\label{eq:v1}
v_\pm = \partial_\pm x\, ,
\end{equation}
while the equations of motion for gauge fields $v_\pm$ give
\begin{equation}\label{eq:vpm1}
v_\pm=\pm \partial_\pm y_1+2H\Delta V\; \partial_\pm y_2\, .
\end{equation}
Inserting expressions for $v_\pm$ into gauge fixed action we get the T-dual action
\begin{equation}
{}_{zyx} S=\kappa \int_\Sigma d^2\xi \partial_+ \;{}_{zyx}X^\mu\; {}_{zyx}\Pi_{+\mu\nu}\;{}_{zyx}\;X^\nu\, ,
\end{equation}
where 
\begin{equation}
{}_{zyx}X^\mu=\left(
\begin{array}{c}
y_1\\
y_2\\
y_3
\end{array}
\right)\, ,\quad  {}_{zyx}\Pi_{+\mu\nu}={}_{zyx}B_{\mu\nu}+\frac{1}{2}\;{}_{zyx}G_{\mu\nu}
\end{equation}
\begin{equation}
{}_{zyx}B_{\mu\nu}=\left(
\begin{array}{ccc}
0 & -H\Delta V & 0\\
H\Delta V & 0 & 0\\
0 & 0 & 0
\end{array}
\right)\, ,\quad
{}_{zyx}G_{\mu\nu}=\left(
\begin{array}{ccc}
1 & 0 & 0\\
0 & 1 & 0\\
0 & 0 & 1
\end{array}
\right)\, .
\end{equation}

Combining the equations of motion (\ref{eq:v1}) and (\ref{eq:vpm1}) we obtain the T-dual transformation law
\begin{equation}\label{eq:tl3}
\partial_\pm x\cong \pm \partial_\pm y_1+2H\Delta V \partial_\pm y_2\, .
\end{equation}
It directly follows that
\begin{equation}\label{eq:dotx}
\dot x\cong y_1'+2H\Delta V \dot y_2\, .
\end{equation}
From the initial action (\ref{eq:dejstvo}) it is obvious that momentum canonically conjugated to $x$ is of the form
\begin{equation}
\pi_x=\kappa \dot x-2\kappa H z y'\, .
\end{equation}
The T-dual transformation law for $y$ (\ref{eq:tl2}), in the approximation linear in $H$, produces that $y'\cong \dot y_2$. Taking into account the relation (\ref{eq:dotx}), we get the canonical form of
the T-dual transformation law
\begin{equation}\label{eq:y1}
y_1'\cong \frac{1}{\kappa}\pi_x\, .
\end{equation}
As we see the full set of T-dual transformation laws, (\ref{eq:y3}), (\ref{eq:y2}) and (\ref{eq:y1}), are the same as in the case where T-dualization was along $xyz$ line \cite{bndo} up to $H\to -H$. 
The full T-dualized theory is of the same form as in \cite{bndo} with the expressions for {\bf noncommutativity} 
\begin{equation}\label{eq:nc1}
\{y_1(\sigma),y_3(\bar\sigma)\}\cong \frac{H}{\kappa}\left[2y(\sigma)-y(\bar\sigma)\right]\theta(\sigma-\bar\sigma)\, ,
\end{equation}
\begin{equation}\label{eq:nc2}
\{y_2(\sigma),y_3(\bar\sigma)\}\cong -\frac{H}{\kappa}\left[2x(\sigma)-x(\bar\sigma)\right]\theta(\sigma-\bar\sigma)\, ,
\end{equation}
and {\bf nonassociativity}
\begin{eqnarray}\label{eq:jacobi}
&& \{y_1(\sigma_1),y_2(\sigma_2),y_3(\sigma_3)\}\equiv \nonumber \\
&&\{y_1(\sigma_1),\{y_2(\sigma_2),y_3(\sigma_3)\}\}+\{y_2(\sigma_2),\{y_3(\sigma_3),y_1(\sigma_1)\}\}+\{y_3(\sigma_3),\{y_1(\sigma_1),y_2(\sigma_2)\}\}\cong\nonumber \\
&&\frac{2H}{\kappa^2}\left[\theta(\sigma_1-\sigma_2)\theta(\sigma_2-\sigma_3)+\theta(\sigma_2-\sigma_1)\theta(\sigma_1-\sigma_3)+\theta(\sigma_1-\sigma_3)\theta(\sigma_3-\sigma_2)\right]\, ,
\end{eqnarray}
which can be obtained from the corresponding ones in $xyz$ case \cite{bndo} by replacing $H\to -H$.

\section{Quantum aspects of T-dualization in the weakly curved background}
\setcounter{equation}{0}

In proving isometry and computing the $\beta^\pm$ functions we assumed the trivial topology and
the surface term occurring there vanishes. Now we want to discuss some quantum as-
pects of the considered problems in nontrivial topologies. We will consider the action for
bosonic string in the weakly curved background - constant metric and Kalb-Ramond field
depending on all coordinates and with infinitesimal field strength. Torus with infinitesimal
$H$-flux is special case of this model.

On th classical level there are a few problems in the theory. In order to perform the
generalized T-dualization procedure the invariant coordinate $x^\mu_{inv}$ is introduced. But it
is multivalued and the proof of equivalence of gauged and initial theories needs the part
considering global characteristics. Moreover, in the quantum theory at higher genus, the holonomies
of the world-sheet gauge fields complicate the situation a little bit. Fortunately, these problems can be
resolved in Abelian case in the quantum theory \cite{RV,AABL,RP}.

First, we make Wick rotation $\tau \to -i \tau$, which makes the term which contains metric
tensor $G_{\mu \nu}$ gets multiplier
$i$, while the terms which contain Kalb-Ramond field $B_{\mu \nu}$
and Lagrange multiplier $y_\mu$ stay unchanged.
Then the partition function is of
the form
\begin{equation}\label{eq:zdef}
Z=\sum_{g=0}^{\infty}\int {\cal D}y {\cal D}v\,
e^{-\frac{\kappa}{2}\int_{\Sigma}  v \,G\, {}^\star v +i\kappa \int_{\Sigma}  v B[V] v
+\frac{i\kappa}{2}\int_{\Sigma}vdy} \,  .
\end{equation}
We use differential forms and omit the space-time indices to simplify writing of equations. The Hodge duality operator is
denoted by star. The index $g$ denotes the genus of manifold..

The first step in the calculation process is separation the one form $dy$ into
the exact part $dy_{e}$ ($y_{e}$ is single valued) and the
 harmonic part $y_{h}$ ($dy_{h}=0=d^\dag
y_{h}$)
\begin{equation}\label{eq:ydec}
dy=dy_{e}+y_{h}.
\end{equation}
For the closed forms the co-exact term $d^\dag y_{co}$
in the Hodge decomposition is missing.

The path integral (\ref{eq:zdef}) goes over all degrees of freedom including local degrees
of freedom as well as the sum over different topologies. Consequently, according to the
(\ref{eq:ydec}), we substitute
${\cal D}y$ with the path integral over $y_{e}$ and the sum
over all possible topologically nontrivial states contained in $y_{h}$
(marked by $H_{y}$)
\begin{equation}
{\cal D}y\rightarrow {\cal D}y_{e}\sum_{H_{y}}.
\end{equation}
The integration over $y_{e}$ induces vanishing of the field strength
\begin{equation}\label{eq:zhol}
Z=\int {\cal D}v\,
\delta(dv)\,
e^{-\frac{\kappa}{2}\int_{\Sigma}  v \, G \, {}^\star v +i\kappa \int_{\Sigma}  v B[V] v}
\sum_{H_{y}}
e^{
\frac{i\kappa}{2}\int_{\Sigma}v y_h}.
\end{equation}

The 1-form $v$ can be expressed as sum of exact, co-exact and the harmonic parts
\begin{equation}
v=dx+d^\dag v_{ce}+v_{h},
\end{equation}
which means that
\begin{equation}
{\cal D}v\rightarrow
{\cal D}x
{\cal D}d^\dag v_{ce}\, dH_{v}.
\end{equation}
The functional integration over harmonic part $v_h$ drives to the ordinary integration over topologically nontrivial periods
(marked by symbol $H_{v}$).
After integration over $d^\dag v_{ce}$ we get
\begin{equation}\label{eq:zholl}
Z=\int {\cal D}x dH_{v}\,
e^{-\frac{\kappa}{2}\int_{\Sigma}  v \, G \, {}^\star v +i\kappa \int_{\Sigma}  v B[V] v}
\sum_{H_{y}}
e^{\frac{i\kappa}{2}\int_{\Sigma}v y_h}.
\end{equation}

The last term in the exponent is responsible for nontrivial holonomies.
Eliminating $v_{ce}$ part,
the 1-form $v$ becomes closed and
the Riemann bilinear relation becomes usable
\begin{equation}
\int_{\Sigma}vy_{h}= \sum_{i=1}^{g}\Big{[} \oint_{a_{i}}v
\oint_{b_{i}}y_{h} -\oint_{a_{i}}y_{h}
\oint_{b_{i}}v \Big{]}.
\end{equation}
The symbols $a_{i},b_{i}\,(i=1,2,\dots,g)$ represent the canonical homology basis
for the world-sheet.
Because of the periodicity of the Lagrange multiplier $y$, its periods are just the winding numbers around cycles $a_{i}$ and $b_{i}$
\begin{equation}\label{eq:nn}
N_{a_{i}}=\oint_{a_{i}}y_{h} ,\quad
N_{b_{i}}=\oint_{b_{i}}y_{h} \, .
\end{equation}
Denoting the periods with
\begin{equation}\label{eq:AB}
A_{i}=\oint_{a_{i}}v,\quad
B_{i}=\oint_{b_{i}}v \,  ,
\end{equation}
we get
\begin{equation}
\int_{\Sigma}v y_{h}=\sum_{i=1}^{g}(N_{b_{i}}A_{i}-N_{a_{i}}B_{i}).
\end{equation}
Now the partition function (\ref{eq:zholl}) gets the form
\begin{equation}\label{eq:zholexp}
Z=\int {\cal D}x\,
dA_{i}dB_{i}
e^{-\frac{\kappa}{2}\int_{\Sigma}  v \, G \, {}^\star v +i\kappa \int_{\Sigma}  v B[V] v}
\sum_{N_{a_{i}},N_{b_{i}}\in Z}
e^{\frac{i\kappa}{2}\sum_{i=1}^{g}(N_{b_{i}}A_{i}-N_{a_{i}}B_{i})}.
\end{equation}
The periodic delta function is defined as
$\delta(x)=\frac{1}{2\pi}\sum_{n\in Z}e^{in x}$, which produces
\begin{equation}\label{eq:zfinal}
Z=\int
{\cal D}x\,
dA_{i}dB_{i}
\delta(\frac{\kappa}{2}A_{i})
\delta(\frac{\kappa}{2}B_{i})\,
e^{-\frac{\kappa}{2}\int_{\Sigma}  v \, G \, {}^\star v +i\kappa \int_{\Sigma}  v B[V] v} \,  .
\end{equation}

It is useful to examine the path dependence of the variable  $V^\mu$, which form is now
\begin{equation}
V^\mu(\xi)=x^\mu(\xi)-x^\mu(\xi_{0})+\int_{P}v_{h}^\mu.
\end{equation}
Let us consider two paths, $P_1$ and  $P_{2}$, with the same initial $\xi^{\alpha}_{0}$
and the final points $\xi^{\alpha}$. Now we will subtract from the value of $V^\mu$ along $P_1$ 
the value along path $P_2$ and obtain
the integral over closed curve $P_1P_{2}^{-1}$ of the
harmonic form
\begin{equation}
V^\mu[P_1]-V^\mu[P_{2}]
=\oint_{P_1P^{-1}_{2}}v_{h}^\mu.
\end{equation}
Establishing the homology between the closed curve $P_1P^{-1}_{2}$ and curve
$\sum_{i}\big{[}n_{i}a_{i}+m_{i}b_{i}\big{]}$, $(n_{i},m_{i}\in Z)$
we get finally
\begin{equation}\label{eq:Vpath}
V^\mu[P_1]=V^\mu[P_{2}]+\sum_{i}(n_{i}A^\mu_{i}+m_{i}B^\mu_{i}).
\end{equation}
The variable $V^\mu(\xi)$ in classical theory is path dependent if holonomies are nontrivial.

Integrating Eq.(\ref{eq:zfinal}) over $A_{i}$ and $B_{i}$ implies that periods $A_{i}$ and $B_{i}$ are zero. Consequently
\begin{equation}
v=dx.
\end{equation}
The variable $V^\mu$ becomes single
valued, and the initial theory is restored
\begin{equation}
Z=\int {\cal D}x
e^{-\frac{\kappa}{2}\int_{\Sigma}  dx \, G \, {}^\star dx +i\kappa \int_{\Sigma}  dx B[x] dx}
=\int {\cal D}x
e^{-\kappa\int_{\Sigma}d^2 \xi \partial x\Pi_{+}[x]{\bar \partial} x} \,  .
\end{equation}

Consequently, starting with partition function of the gauged fixed action of bosonic string in the weakly curved background, within path integral formalism and in the presence of nontrivial topologies, we came to the partition function of the initial theory. That means that introducing coordinate dependent Kalb-Ramond field is consistent with path integral quantization process.

\section{Conclusion}
\setcounter{equation}{0}

In this article we studied the 3D closed bosonic string propagating in the geometry known as torus with $H$-flux - constant metric and Kalb-Ramond field with just one nonzero component, $B_{xy}=-B_{yx}=Hz$. 
The choice of background fields is consistent with the consistency conditions if we work in the diluted flux approximation which assumes that in all calculations we keep just the constant terms and those linear
in the infinitesimal field strength $H$. Our goal was to study the T-dualization line which goes in the opposite direction from the standard one. First, we T-dualize $z$ direction, then $y$ and at the end along
$x$ direction - so-called $zyx$ T-dualization line. We analyzed in every step the (non)commutativity and (non)associativity of the obtained theory and made comparisons with the case of $xyz$ T-dualization line considered
in \cite{bndo,ALLP}.

The common fact for all three theories obtained in the process of T-dualization step by step is that all three ones are nonlocal R-flux theories. The nonlocality comes as a result of the first step in T-dualization
procedure, T-dualization along $z$ direction. Generalized T-dualization procedure has one additional step with respect to the standard Buscher procedure and that is introduction of invariant coordinate. In the process
of T-dualization invariant coordinate turns into variable $\Delta V$ which is defined as line integral. Consequently, this means that obtained theory is nonlocal. Further T-dualizations does not affect $\Delta V$
and, all three theories are nonlocal ones. As we know, in the case of $xyz$ T-dualization line \cite{bndo}, we obtained three different theories in geometrical sense - twisted torus, $Q$-flux theory (which is local) and 
nonlocal $R$-flux theory.

The dualization along $z$ direction produces nonlocal R-flux theory unlike the $xyz$ case \cite{bndo,ALLP} where the theory obtained after first T-dualization is locally and globally well defined. Because initial theory
is geometrical one, its variables satisfy standard Poisson algebra (\ref{eq:spa}). Using (\ref{eq:spa}) and T-dual transformation law written in the canonical form (\ref{eq:y3}), we showed that theory obtained
after T-dualization along $z$ coordinate (using generalized T-dualization procedure) is {\it commutative}
and, consequently, {\it associative} one as in \cite{bndo}.

The second step in T-dualization is T-dualization along $y$ direction. Using standard Buscher procedure, we obtained the form of the T-dual theory and the corresponding T-dual transformation law, which is rewritten 
in the canonical form (\ref{eq:y2}) in terms of the coordinates and momenta of the initial theory. Using standard Poisson algebra (\ref{eq:spa}) and T-dual transformation laws in canonical form, 
(\ref{eq:y3}) and (\ref{eq:y2}), we easily proved that theory after two T-dualizations is {\it noncommutative}, but it is still {\it associative} one. In this article we used trivial winding condition (\ref{eq:winding}) and showed that T-dual coordinates $y_2(\sigma)$ and $y_3(\bar\sigma)$
are commutative for equal arguments, $\sigma=\bar\sigma$, but they are noncommutative if $\sigma-\bar\sigma=2\pi$. The result is qualitatively similar to the result of \cite{ALLP}, where after two T-dualizations the 
obtained theory is noncommutative one. But, the difference is in the winding condition which is nontrivial in \cite{ALLP}, mixing different coordinates. The different winding condition induces the noncommutativity for
$\sigma=\bar\sigma$ (for more details see \cite{ALLP}). On the other hand in the analysis presented in \cite{bndo} ($xyz$ T-dualization line) the theory obtained after two T-dualizations is commutative under trivial winding
condition.

The final step in T-dualization procedure is T-dualization along $x$ direction. The theory after full T-dualization is the same as in $xyz$ case \cite{bndo} with the noncommutativity and nonassociativity
parameters which can be obtained from those in $xyz$ case \cite{bndo} adding "$-$" sign. This is a consequence of the fact that the full set of T-dual transformation laws is the same as in \cite{bndo} up to the 
replacing $H\to - H$. This difference up to the "$-$" sign stems from the initial actions. In this article we start from (\ref{eq:dejstvo}), while in \cite{bndo} the starting action for $z$ T-dualization is
$Q$-flux action, formally the same as (\ref{eq:dejstvo}) up to the replacing $H\to -H$.

Finishing the discussion of the results obtained in this paper it is inter-
esting to make comparison with some similar efforts. We studied the abelian
isometries using both standard and generalized T-duality procedure, while in
the paper \cite{ossa} nonabelian isometries using standard Buscher procedure are
considered. The authors of \cite{ossa} showed that spaces with isometry maps to
the nonisometry spaces, while in this paper there is isometry in every T-
dualization step. One of their conclusions that T-dual transformations are
more than continuous isometry can be added to the concluding remarks of
this paper. In the Ref.\cite{nongeo2} generalized T-duality and nongeometric background are considered, but using low energy effective action,
unlike here, where we used sigma model action. The paper \cite{acadlj} deals with T-dualizations along nonisometry directions like in \cite{DNS3}, using extension of gauge
symmetry, while the authors of \cite{DNS3} use the generalized T-dualization procedure introducing invariant coordinates (in \cite{acadlj} they call them ”covariant”
coordinates). In this paper we use this generalized T-dualization procedure
but all directions considered here are isometry ones. It is useful to mention that in the paper \cite{bsdouble} bosonic string in the presence of the weakly curved backgrounds is considered using double space formalism as well as the influence of the order of T-dualizations. The double space formalism gives the result which is in accordance with the result of the current paper.

Consequently, we conclude that in the case of the full T-dualization the form of the T-dual
theory do not depend on the order of T-dualization, while parameters of noncommutativity and nonassociativity change sign.

\appendix
\setcounter{equation}{0}

\section{Light-cone coordinates}

In the paper we often use light-cone coordinates defined as
\begin{equation}
\xi^\pm=\frac{1}{2}(\tau\pm\sigma)\, .
\end{equation}
The corresponding partial derivatives are
\begin{equation}
\partial_\pm\equiv \frac{\partial}{\partial \xi^\pm}=\partial_\tau \pm \partial_\sigma\, .
\end{equation}

Two dimensional Levi-Civita $\varepsilon^{\alpha\beta}$ is chosen in $(\tau,\sigma)$ basis as $\varepsilon^{\tau\sigma}=-1$. Consequently, in the light-cone basis the form of tensor is
\begin{equation}
\varepsilon^{\alpha\beta}_{lc}=\left(
\begin{array}{cc}
0 & \frac{1}{2}\\
-\frac{1}{2} & 0
\end{array}\right)\, .
\end{equation}
The flat world-sheet metric is of the form in $(\tau,\sigma)$ and light-cone basis, respectively
\begin{equation}
\eta_{\alpha\beta}=\left(
\begin{array}{cc}
1 & 0\\
0 & -1
\end{array}\right)\, ,\quad \eta^{lc}_{\alpha\beta}=\left(
\begin{array}{cc}
\frac{1}{2} & 0\\
0 & \frac{1}{2}
\end{array}\right)\, .
\end{equation}
Let us stress that in whole article we use standard notation for $\tau$ and $\sigma$ derivatives - $\dot A\equiv \partial_\tau A$ and $A'\equiv \partial_\sigma A$, where $A$ is an arbitrary variable.

\section{Two types of Poisson brackets used in the paper}

In this paper, we have seen that T-dual transformation laws connect derivatives of T-dual coordinates with coordinates and momenta of initial theory. While initial theory satisfies standard Poisson brackets, in order to find Poisson brackets for T-dual theory, we first need to find Poisson brackets between $\sigma$ derivatives of T-dual coordinates. This type of Poisson bracket will, in general case, be some function of initial coordinates, Dirac delta functions and their derivatives with respect to $\sigma$. Having this in mind, general case for our Poisson brackets will have following form
\begin{equation}\label{eq:sigmapoisson}
\{A'(\sigma),B'(\bar\sigma)\}=U'(\sigma)\delta(\sigma-\bar\sigma)+V(\sigma)\delta'(\sigma-\bar\sigma)\, ,
\end{equation}
where $\delta'(\sigma-\bar\sigma)\equiv \partial_\sigma \delta(\sigma-\bar\sigma)$. 
For terms $A'(\sigma)$, $U'(\sigma)$ and $B'(\bar{\sigma})$, symbol $'$ stands for partial derivative with respect to $\sigma$ and $\bar\sigma$, respectively.
If we want to calculate the Poisson bracket
$$\{A(\sigma),B(\bar\sigma)\}\, ,$$
first we have to calculate the following one
\begin{equation}\label{eq:PBD}
\{\Delta A(\sigma,\sigma_0),\Delta B(\bar\sigma,\bar\sigma_0)\}\, ,
\end{equation}
where
\begin{equation}\label{eq:Delte}
\Delta A(\sigma,\sigma_0)=\int_{\sigma_0}^\sigma dx A'(x)=A(\sigma)-A(\sigma_0)\, ,\quad \Delta B(\bar\sigma,\bar\sigma_0)=\int_{\bar\sigma_0}^{\bar\sigma} dx B'(x)=B(\bar\sigma)-B(\bar\sigma_0)\, .
\end{equation}
Substituting the expressions (\ref{eq:Delte}) into (\ref{eq:PBD}), we have
\begin{equation}
\{\Delta A(\sigma,\sigma_0),\Delta B(\bar\sigma,\bar\sigma_0)\}=\int_{\sigma_0}^\sigma dx \int_{\bar\sigma_0}^{\bar\sigma}dy\;\left[U'(x)\delta(x-y)+V(x)\delta'(x-y)\right]\, .
\end{equation}
After integration over $y$ we get
\begin{eqnarray}
&{}&\{\Delta A(\sigma,\sigma_0),\Delta B(\bar\sigma,\bar\sigma_0)\}=\nonumber\\
&=&\int_{\sigma_0}^\sigma dx \{U'(x)\left[\theta(x-\bar\sigma_0)-\theta(x-\bar\sigma)\right]+V(x)\left[\delta(x-\bar\sigma_0)-\delta(x-\bar\sigma)\right]\},
\end{eqnarray}
where $\theta(x)$ is defined as
\begin{equation}\label{eq:fdelt}
\theta(x)=\int_0^x d\eta\delta(\eta)=\frac{1}{2\pi}\left[x+2\sum_{n\ge 1}\frac{1}{n}\sin(nx)\right]=\left\{\begin{array}{ll}
0 & \textrm{if $x=0$}\\
1/2 & \textrm{if $0<x<2\pi$}\, ,\\
1 & \textrm{if $x=2\pi$} \end{array}\right .
\end{equation}
{\bf where $\delta(x)=\frac{1}{2\pi}\sum_{n\in Z} e^{inx}$}.
Finally, integrating over $x$, we obtain
\begin{eqnarray}
&&\{\Delta A(\sigma,\sigma_0),\Delta B (\bar{\sigma},\bar{\sigma}_0)  \} = \nonumber\\
&&U(\sigma)[\theta(\sigma-\bar{\sigma}_0)-\theta(\sigma-\bar{\sigma})  ]-U(\sigma_0)[\theta(\sigma_0-\bar{\sigma}_0)-\theta(\sigma_0-\bar{\sigma})  ]\nonumber\\
&-&U(\bar{\sigma}_0)[\theta(\sigma-\bar{\sigma}_0)-\theta(\sigma_0-\bar{\sigma}_0) ]+U(\bar{\sigma})[\theta(\sigma-\bar{\sigma})-\theta(\sigma_0-\bar{\sigma})  ]\nonumber\\
&+&V(\bar{\sigma}_0)[\theta(\sigma-\bar{\sigma}_0)-\theta(\sigma_0-\bar{\sigma}_0  ]-V(\bar{\sigma})[\theta(\sigma-\bar{\sigma})-\theta(\sigma_0-\bar{\sigma})].
\end{eqnarray}
From the last expression, using (\ref{eq:Delte}), we extract the searched Poisson bracket
\begin{equation}\label{eq:tdpoisson}
\boxed{\{A(\sigma),B(\bar\sigma)\} =-[U(\sigma)-U(\bar{\sigma})+V(\bar{\sigma})  ]\theta(\sigma-\bar{\sigma})}\, .
\end{equation}

In order to calculate Jacobiator we have to find Poisson brackets of type $\{y(\sigma),x(\bar\sigma)\}$, where $y(\sigma)$ is coordinate T-dual to initial one $x(\sigma)$. Having this in mind, we start with the following Poisson bracket
\begin{equation}
\{\Delta y(\sigma,\sigma_0),x(\bar\sigma)\}=\{\int_{\sigma_0}^\sigma d\eta y'(\eta),x(\bar\sigma)\}\, ,
\end{equation}
and using T-dual transformation law in canonical form
\begin{equation}
\pi\cong \kappa y'\, ,
\end{equation}
we get
\begin{equation}
\{\Delta y(\sigma,\sigma_0),x(\bar\sigma)\}\cong \frac{1}{\kappa} \{\int_{\sigma_0}^\sigma d\eta \pi(\eta), x(\bar\sigma)\}\, ,
\end{equation}
where $\pi(\sigma)$ is momentum canonically conjugated to the coordinate $x(\sigma)$.
Initial theory is geometric one which variables satisfy standard Poisson algebra, so, the final result is of the form
\begin{equation}
\{\Delta y(\sigma,\sigma_0),x(\bar\sigma)\}\cong -\frac{1}{\kappa}\left[\theta(\sigma-\bar\sigma)-\theta(\sigma_0-\bar\sigma)\right] \quad \Longrightarrow \boxed{\{y(\sigma),x(\bar\sigma)\}\cong -\frac{1}{\kappa}\theta(\sigma-\bar\sigma)}\, .
\end{equation}



\end{document}